\documentclass[aps,prb,twocolumn,nopacs,floats,superscriptaddress]{revtex4-2}
\usepackage{graphicx}
\usepackage{dcolumn}
\usepackage{bm}
\usepackage{amsmath}
\usepackage{color}

\definecolor{cardinal}{rgb}{0.6,0,0}
\definecolor{darkgreen}{rgb}{0,0.4,0}
\definecolor{golden}{rgb}{0.92, 0.7, 0}
\definecolor{midnight}{rgb}{0, 0, 0.5}
\definecolor{darkblue}{rgb}{0, 0, 0.7}

\def\he4{$^4$He}
\def\hel3{$^3$He}
\def\Am3{\AA$^{-3}$}
\def\beq{\begin{equation}}
\def\eeq{\end{equation}}

\newcommand{\be}{\begin{equation}}
\newcommand{\ee}{\end{equation}}
\newcommand{\bea}{\begin{eqnarray}}
\newcommand{\eea}{\end{eqnarray}}
\newcommand{\bse}{\begin{subequations}}
\newcommand{\ese}{\end{subequations}}

\begin{document}
\author{Adham Alkady({\textdagger})}



\author{Anatoly Kuklov}
\affiliation{Department of Physics \& Astronomy, College of Staten Island and the Graduate Center of
CUNY, Staten Island, NY 10314}


\title{Symmetry induced pairing in dark excitonic condensate at finite temperature.}

\begin{abstract}
 Bose Einstein condensate of dark intervalley excitons must be inherently multi-component because of crystalline symmetries. Since valleys hosting such excitons are separated by large quasi-momenta, a minimal inter-component Josephson-type coupling can only be established between pairs of excitons from the time-reversed valleys.  As a result, a paired condensate can emerge at finite temperature, that is, the off-diagonal order exists for the pairs from the time-reversed valleys, while the individual valleys are disordered. This prediction follows from the elementary mean field analysis regardless of the dimensionality. However, as  Monte Carlo simulations show, no such a phase exists in 3D crystals. Instead, the multi-component condensation proceeds as the Ist order transition from the normal state. The paired phase does exist in 2D for the number of the components $N_v\geq 6$. It forms by Berezinskii-Kosterlitz-Thouless transition from the high temperature (normal) phase. The multi-component condensate appears upon further lowering of temperature.
\end{abstract}

\maketitle

\section{Introduction}

In contrast to the bright excitons, dark excitons cannot be created directly by light (see in Ref.\cite{BEDE}). A typical example, Ref.\cite{Gamma}, is a momentum forbidden (intervalley) exciton formed by a hole at the $\Gamma-$point and an electron at a valley characterized by some momentum $\vec{Q}$ significantly larger than a typical momentum of light.
The resulting exciton, then, is optically inactive because it carries the center-of-mass momentum  $\vec{Q}$.
Accordingly, dark excitons can normally persist much longer than the bright ones \cite{long}, and this creates an attractive perspective for realizing  collective states of the excitonic ensembles such as  Bose-Einstein condensates \cite{Keldysh,Moskalenko}, the Mott-insulator phases \cite{Mott} and the topological order \cite{Sedrakyan}. 

As will be discussed below, the condensate of such excitons adds to the family of multi-component superfluids exhibiting such properties as fractional vortices \cite{Egor} paired phases \cite{Egor2}, formation of quasi-molecular complexes \cite{Kaurov}, knot solitons \cite{knot} and counter-flow phases \cite{SCF}  (see also in Ref. \cite{book}).   While earlier studies of the multi-component systems focused on mostly two-component cases, recently it was pointed out in Refs.\cite{borom1,borom} that the systems with three and more components set the stage for essentially new qualities---the so called Borromean counter-superfluids.

 For the intervalley dark-exciton ensembles, a special role is played by crystal symmetry. Focus of this paper is on the interplay between this symmetry and possible excitonic phases.

\section{The role of crystal symmetry} 
In contrast to the traditional condensation at zero momentum, the intervalley exitonic condensate must occur at the valley momentum, so that the corresponding bosonic field \cite{LL} can be represented as $\Psi(\vec{x})=\psi(\vec{x}) \exp(i \vec{Q}\vec{x})$, where $\psi(\vec{x})=\sqrt{n}\exp(i\varphi)$ stands for the amplitude determined by the density $n$ and the collective phase  $\varphi$ of the condensed excitons. Accordingly, the density matrix describing the off-diagonal order \cite{LL}  $\rho(\vec{x}, \vec{y}) = \langle \psi^*(\vec{y}) \psi(\vec{x})\rangle \exp(i\vec{Q}(\vec{x}-\vec{y}))$ contains the fast oscillating part even if the correlator
$\langle \psi^*(\vec{y}) \psi(\vec{x})\rangle$ is long ranged.
The kinetic part of the Hamiltonian for a $\nu$-th valley can be represented as 
\beq
H_\nu =\int d^d x \frac{\hbar^2}{2m} |(\vec{\nabla} -i \vec{Q}_\nu) \psi_\nu |^2,
\label{kin}
\eeq
where $m$ stands for the effective mass at the bottom of the valley.
 This situation is equivalent to the case of the traditional condensate described from the frame moving at the speed $ \hbar \vec{Q}_\nu/m$. Obviously, in this case the oscillating part is absolutely inconsequential, and from all perspectives the oscillating part can be ignored, that is, the standard Gross-Pitaevskii  description \cite{LL} can be applied to the order parameter $\psi_\nu$. 

\begin{figure}[!htb]
\includegraphics[width=0.75 \columnwidth]{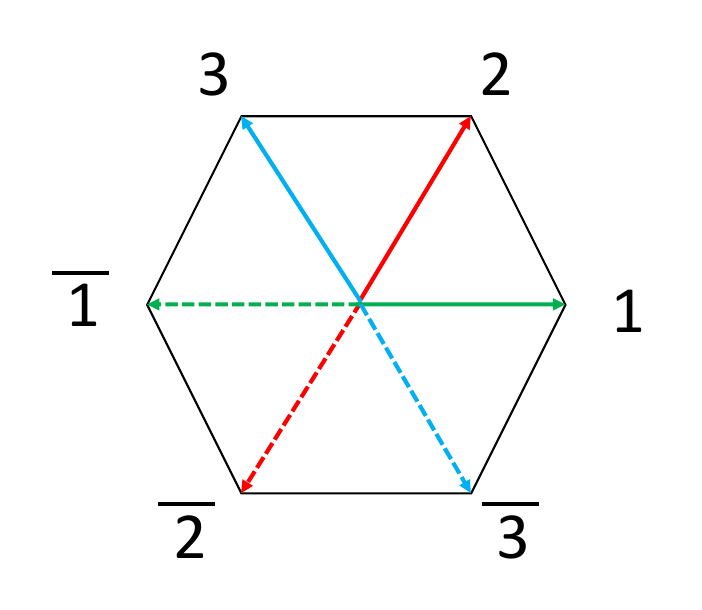}
\caption{ A sketch of the valley positions $\vec{Q}_\nu$ in the momentum space of a hexagonal crystal. The bar indicates the time-reversed valleys, so that, if counting counter clockwise, the valley 4 is the time-reversed partner to the valley 1, and, thus, is shown as $\bar{1}$. Similarly,---for the pairs 2,5 (as $\bar{2}$) and 3,6 (as $\bar{3}$).  }
\label{fig0}
\end{figure}

The point-group of a crystal demands that there are several valleys characterized by the same excitonic energy. Accordingly, the relaxation of the dark excitons should proceed into a multi-component condensate, with each component  characterized by the field $\Psi_\nu = \psi_\nu e^{i \vec{Q}_\nu \vec{x}}$, where $\nu=1,2,..., N_v$ and $N_v$ is the total number of valleys  representing crystal symmetry.   In the case of spinless excitons, the time-reversal symmetry demands that for each valley $\nu$ there is its time-reversed partner $\bar{\nu}$ so that $\vec{Q}_{\bar{\nu}} =- \vec{Q}_{\nu}$ and all values $|\vec{Q}_\nu|$ are the same. If the spin-orbit coupling is present \cite{SO}, the same relations hold if spin reversal is taken into account.  
 A typical  example of such a system is a hexagonal layer of TMD material \cite{BEDE,Gamma} characterized by $N_v=6$ as depicted in Fig.~\ref{fig0}. 

The interaction between the components can be written as $ \int d^2x [g_1 \sum_{\mu} |\psi_\mu|^4 +  g_2\sum_{\mu > \nu}  |\psi_\mu|^2  |\psi_\nu|^2]$ with some constants $g_1$ and $g_2$ (both considered as positive). Stability against phase separation requires that $g_2\leq 2g_1$, with the equality corresponding to the  symmetry U(N$_v$) as long as there is no internal Josephson coupling between the components. If $g_2<2g_1$, the symmetry is lowered to [U(1)]$^{N_v}$. [Here the possibility of the U(N$_v$) symmetry is not considered].   

Excitons from different valleys can be coupled by a Josephson-type term. However, the traditional coupling $\sim \int d^dx \Psi^*_\nu \Psi_{\nu'} +c.c. \sim \int d^dx \cos( (\vec{Q}_\nu -\vec{Q}_{\nu'})\vec{x})$ is strongly suppressed because of the large momentum mismatch $|\vec{Q}_\nu - \vec{Q}_{\nu'}|$ between valleys. 
Thus, valleys host $N_v$ independent condensates in the lowest order with respect to the Josephson coupling. This case represents the opposite limit to the situation when $N_v\to \infty$ and the valleys merge into a moat \cite{Sedrakyan} where the condensate can transform into a spin-liquid type phase \cite{Sedrakyan1,Sedrakyan2,Sedrakyan3}.   

Valleys can be coupled by the second order Josephson effect transforming excitons from one pair $(\nu, \bar{\nu})$ of the time reversed valleys into another pair, ($\mu,\bar{\mu}$) \cite{Adham}. These processes are described by the term 
\beq
H_2 = -\frac{g}{2} \int d^2x \sum_{\mu \neq \nu}  (\Psi^*_{\nu} \Psi^*_{\bar{\nu}} \Psi_{\mu} \Psi_{\bar{\mu}} + c.c.), 
\label{HJ1}
\eeq
 where $g>0$ is some parameter. Counting phases admitting independent global U(1) transformations shows, this term lowers the symmetry from [U(1)]$^{N_v}$ to [U(1)]$^{1+N_v/2}$.  A second order internal Josephson coupling between excitons was introduced in Ref.~\cite{Combescot1} for the interconversion between bright and dark (spin-forbidden) excitons. Such coupling leads to specific Rabi oscillations in the luminescence. 

 The main finding reported here is that the term $H_2$ generates pairing of the condensates in 2D, so that for $N_v\geq 6$ lowering temperature $T$ from the normal phase the condensation occurs into the paired state where the corresponding order parameter is $\langle \Psi_\nu \Psi_{\bar{\nu}}\rangle=\langle \psi_\nu \psi_{\bar{\nu}}\rangle  \sim e^{i \Phi}$ with the same phase $\Phi$ for all pairs $\nu, \bar{\nu}$, while each individual field amplitudes $\psi_\nu$ are disordered. 
It is important to realize that the field $\tilde{\Psi}  =\Psi_\nu \Psi_{\bar{\nu}}$ does not contain the fast oscillating part, that is, each pair has zero momentum. This situation corresponds to spontaneously breaking the symmetry [U(1)]$^{1+N_v/2}$ of the normal phase to [U(1)]$^{N_v/2}$. Lowering $T$ further leads to another transition---into the phase where all $\psi_\nu$ are ordered implying that the remaining symmetry [U(1)]$^{N_v/2}$ is fully broken.  In 3D the situation is very different: upon lowering $T$ , the system bypasses the paired phase and enters the $N_v$-component condensate by Ist order phase transition.

 A minimal model of $N_v$ condensates coupled by the term $H_2$ (\ref{HJ1}) is addressed analytically and numerically using Monte Carlo.  The relevant values of $N_v$ consistent with the crystal symmetries are $N_v=2,4,6$. However, in order to get a deeper insight into the mechanism, arbitrary values of $N_v$ are considered as an independent parameter. In principle, quasi-2D situations with N$_v=12,18, ...$ can be realized in twisted bi-,  tri-, etc.-layers of TMDs, respectively. 
  
\section{The minimal model}\label{SEC1}

The minimal description is based on the XY-lattice model for $N_v$ condensates described by  the fields $\psi_\nu = \exp(i \varphi_\nu)$, where the fluctuations of the amplitudes are ignored. Then, the Hamiltonian becomes
\bea
&&H= - \sum_{<ij>,\nu} t \cos(\nabla_{\hat{\alpha}} \varphi_{\nu}) + H_2,
\label{Htot} \\
&&H_2 = - g\sum_{i, \nu \neq \mu} \cos(\varphi_{i,\nu}+\varphi_{i,\bar{\nu}} - \varphi_{i,\mu}-\varphi_{i,\bar{\mu}}) ,
\label{H_J}
\eea
where the summation $\sum_{<ij>}$ is performed over  nearest neighbors sites $i,j=i+\hat{\alpha}$, with $<ij>$ denoting links between such sites and $\hat{\alpha}$ indicating unit vector along a positive link direction;  $\nabla_{\hat{\alpha}} \varphi_{\nu}\equiv \varphi_{i,\nu} - \varphi_{i+\hat{\alpha},\nu}$ stands for the lattice gradient ; $t>0$ is a parameter. The summation in $\nu$ in the first term runs from $\nu=1$ to $\nu=N_v$, and in the second from $\mu,\nu =1$ to $\mu,\nu = N_v/2$.
 Formally, the term (\ref{Htot}) should contain the "gauge" valley contribution $\vec{Q}_\nu$ as in Eq.(\ref{kin}).   
However, in the case when no linear coupling between valleys exists, this term can be removed by the Galilean transformation.

This is a classical model where quantum fluctuations are ignored. Accordingly, temperature $T$ is absorbed into the definition of the parameters $t,g$, and  $t^{-1}$ can be treated as temperature. 
The periodic boundary conditions for $\varphi_{i,\nu}$ are used. 
The goal is evaluating the partition function $Z=\int D\varphi_{i,\nu} \exp(-H)$. Before, however, it is instructive to conduct an elementary analysis hinting on the possibility of the paired phase. 
Let's introduce the variables $\varphi_{i,\nu} = (\Phi_i +\phi_{i,\nu})/2, \,\, \varphi_{i,\bar{\nu}} = (\Phi_i -\phi_{i,\nu})/2 $, with $\nu=1,2,...N_v/2$ and choose all $\Phi_i =\Phi$ in order to minimize the part (\ref{H_J}).
In these variables 
\bea
\psi_\nu\psi_{\bar{\nu}}= e^{i\Phi}, \,\, \psi_\nu\psi^*_{\bar{\nu}}=e^{i\phi_\nu},
\label{var} 
\eea
where $\nu=1,...,N_v/2$ and $\bar{\nu}= N_v/2 +1, ..., N_v$, with $\bar{1}=N_v/2 +1,\, \bar{2}= N_v/2 +2,... $ in line with that shown in Fig.~\ref{fig0}. Thus, the pairing phase corresponds to order in $\tilde{\Psi}=e^{i\Phi}$, while all $e^{i\phi_\nu}$ are disordered.
Using the gaussian approximation $\cos(\nabla_{\hat{\alpha}} \varphi_{\nu}) \to  1-(\nabla_{\hat{\alpha}} \varphi_{\nu})^2/2$, the functional (\ref{Htot}) becomes
\bea
H \to \tilde{H}= \sum_{<ij>} \left[\frac{tN_v}{8} (\nabla_{\hat{\alpha}} \Phi)^2 +\sum^{N_v/2}_{\nu=1}   \frac{t}{4}(\nabla_{\hat{\alpha}} \phi_{\nu} )^2\right].
\label{Hgauss}
\eea
Considering $\Phi$ and $\phi_\nu$ as compact independent variables defined modulo $2\pi$, the stiffness $tN_v/8$ of $\Phi$ is the factor $N_v/2$ higher than that $t/4$ of the variables $\phi_\nu$. Accordingly, it is tempting to claim that, if the critical value for the $\phi-$variables is $t_c$ (so that for $t<t_c$ there is no order), the critical value for $\Phi$ must be $\tilde{t}_c =2t_c/N_v$, with the paired phase existing for the temperatures $t^{-1}_c  <t^{-1} <N_vt^{-1}_c/2 $ for $N>2$. However, as the simulations will show, this prediction turns out to be marginal. First,  it is wrong in 3D because no paired phase has been found in simulations, and, instead, the superfluid phase with $\Phi$ and $\phi_\nu$ all ordered, transforms into the normal phase by Ist order transition for $N_v>2$. Second, in 2D the paired phase exists only for $N_v=6,8, ...$ with the critical temperature behaving as $\tilde{t}^{-1}_c \sim \ln N_v$ for large $N_v$ instead of $\tilde{t}^{-1}_c \sim N_v$. 

Another doubt follows from using the variables $\Phi, \phi_\nu$ in Eq.(\ref{Htot}) giving  
\bea
\tilde{H}= - \sum_{<ij>}\left[ 2t \cos(\nabla_{\hat{\alpha}} \Phi/2 )\sum^{N_v/2}_{\nu=1} \cos(\nabla_{\hat{\alpha}} \phi_{\nu}/2 )\right].
\label{H2}
\eea
This form suggests just one transition. Indeed, if $\phi_\nu $ are disordered, the effective stiffness $\tilde{t} \approx 2t\langle  \cos(\nabla_{\hat{\alpha}} \phi_{\nu}/2 )\rangle$ for $\Phi$ , where $\langle ...\rangle$ implies statistical averaging,  tends to zero, and vice verse. 

Such ambiguities indicate that the system is strongly interacting where thermal fluctuations play a central role. Below we present the results of the numerical analysis.

\section{ Monte Carlo simulations}
Phases can be characterized by how stiffnesses  $\rho_{\Phi,\hat{\alpha}}$ and $\rho_{\nu,\hat{\alpha}}$ along the lattice direction $\hat{\alpha}$ of the fields $e^{i\Phi}$ and $e^{i\phi_\nu}$, respectively, evolve in the thermodynamic limit of the system linear size $L\to \infty$. 
These can be defined as second derivatives of the free energy $F(\vec{A}_\nu)$ with respect to artificial vector potential $\vec{A}_\nu \equiv A_{\nu,\hat{\alpha}}$=const introduced into Eq.(\ref{Htot})
as $\cos(\nabla_{\hat{\alpha}} \varphi_{\nu}) \to \cos(\nabla_{\hat{\alpha}} \varphi_{\nu} - A_{\nu,\hat{\alpha}}/L) $ .
It is convenient to redefine the gauge fields in terms of $A_{\Phi,\hat{\alpha}}/L$, $B_{\nu,\hat{\alpha}}/L$ introduced into the form (\ref{H2}) as
\bea
\tilde{H}(A,B)=&& - \sum_{<ij>}\Big[ 2t \cos(\nabla_{\hat{\alpha}} \Phi/2  - A_{\Phi,\hat{\alpha}}/L_{\hat{\alpha}})\cdot
\nonumber \\
&&\sum^{N_v/2}_{\nu=1} \cos(\nabla_{\hat{\alpha}} \phi_{\nu}/2 - B_{\nu,\hat{\alpha}}/L_{\hat{\alpha}})\Big].
\label{H3}
\eea
Writing the cos-product as the sum, one finds
\beq
A_{\nu,\hat{\alpha}} =A_{\Phi,\hat{\alpha}} + B_{\nu,\hat{\alpha}} , \,\,\, A_{\bar{\nu},\hat{\alpha}} = A_{\Phi,\hat{\alpha}} - B_{\nu,\hat{\alpha}}, \,\, 
\label{AB} 
\eeq
where $\nu=1,2,...,N_v/2$.
Then, 
\beq
\rho_{\Phi,\hat{\alpha}} =\frac{1}{L^{d-2}} \frac{\partial^2 F}{\partial  A_{\Phi,\hat{\alpha}}^2} ,\,\, \rho_{\nu,\hat{\alpha}} =\frac{1}{L^{d-2}} \frac{\partial^2 F}{\partial  B_{\nu,\hat{\alpha}}^2} ,
\label{stiff}
\eeq
where $d=2,3$, and the lattices with the same linear sizes along each direction, $L_{\hat{\alpha}}=L$, are considered; 
the limit $A\to 0, B\to 0$ is to be taken after the differentiation; and 
\beq
F=- \ln Z,\,\,Z= \int D\varphi_\nu  D\varphi_{\bar{\nu}} e^{-H(A,B)},
\label{F}
\eeq
where the vector potentials (\ref{AB}) are introduced into (\ref{Htot}) as described above.
So defined stiffnesses characterize each phase. Specifically, in the paired superfluid (PSF) phase $\rho_{\Phi,\hat{\alpha}}$ is finite while $\rho_{\nu,\hat{\alpha}} =0$; in the countersuperfluid (CSF) state \cite{Egor,SCF,borom1}  the situation is reversed---$\rho_{\Phi,\hat{\alpha}}=0$, while $\rho_{\nu,\hat{\alpha}}$ is finite. If the full group of symmetry is broken, that is, in the $N_v$-component superfluid (N$_v$-SF), both stiffnesses become finite.

\subsection{Duality}
It is convenient to implement the Villain representation \cite{Villain} of the model (\ref{Htot},\ref{H_J}) by replacing cos-functions as $\cos(\nabla_{\hat{\alpha}} \varphi_\nu) \to 1-\frac{1}{2}(  \nabla_{\hat{\alpha}} \varphi_\nu +2\pi m_{\nu,ij})^2$ and $\cos(\varphi_{i,\nu}+\varphi_{i,\bar{\nu}} - \varphi_{i,\mu}-\varphi_{i,\bar{\mu}}) \to
 1-\frac{1}{2}(\varphi_{i,\nu}+\varphi_{i,\bar{\nu}} - \varphi_{i,\mu}-\varphi_{i,\bar{\mu}}   +2\pi p_{\mu,\nu; i})^2$, with $p_{\mu,\mu;i}=0$,
on the expense of introducing the integer link variables $m_{\nu,\hat{\alpha}}=0,\pm 1,\pm 2, ...$, and the site integer variables $ p_{\mu,\nu; i}=0,\pm 1, \pm 2, ...$,
with the partition function, Eq.(\ref{F}), transformed as
\beq
Z\to  Z= \sum_{\{m_{\nu,\hat{\alpha}}\}, \{p_{\mu,\nu; i}\}} \int D\varphi_\nu  D\varphi_{\bar{\nu}} e^{-H(A,B)},
\label{F2}
\eeq
where now
\bea
&&H= \sum^{N_v}_{<ij>,\nu=1} \frac{t}{2} (\nabla_{\hat{\alpha}} \varphi_{\nu} - A_{\nu,\hat{\alpha}}/L_{\hat{\alpha}} +2\pi m_{\nu,\hat{\alpha}})^2+
\nonumber \\
 &&\frac{g}{2}\sum_{i, \nu \neq \mu} (\varphi_{i,\nu}+\varphi_{i,\bar{\nu}} - \varphi_{i,\mu}-\varphi_{i,\bar{\mu}}+2\pi p_{\mu,\nu; i})^2 ,
\label{HJ}
\eea
and $A_{\nu,\hat{\alpha}}$ are given in Eq.(\ref{AB}).
Using the Poisson identity $\sum_{n=0,\pm 1, \pm 2,...} f(n) = \sum_{J=0,\pm 1, \pm 2,...} \int dx f(x)\exp(2\pi iJ x)$ on each site and link,
the integrations and summations in the partition function (\ref{HJ}) can be done explicitly, which results in the so called J-current version \cite{Jcurr} of the model   
(\ref{F2}) as
\beq
  Z= \sum_{\{J_{\nu,ij}\}, \{K_{\mu,\nu; i}\}} e^{-H_J +\sum_{<ij>, \nu} i J_{\nu; i,\hat{\alpha}} A_{\nu, \hat{\alpha}}/L_{\hat{\alpha}}},
\label{F3}
\eeq
where $ J_{\nu; i, \hat{\alpha}}=0,\pm 1, \pm 2,...$ is the integer link current assigned to the link between sites $i, j=i+\hat{\alpha}$ and   
\beq
  H_J= \sum_{\nu,<ij>} \frac{J^2_{\nu;i,\hat{\alpha}}}{2t} + \sum_{\mu\neq \nu, i} \frac{K^2_{\mu,\nu;i}}{2g},
\label{HJ2}
\eeq
with $ J_{\nu; i,\hat{\alpha}}$ and the site variables $K_{\mu,\nu;i}=- K_{\nu,\mu;i}$ satisfying the following constraints on each site (site index is not shown):
\bea
&&\vec{\nabla} \vec{J}_\nu+ \sum_{\mu} K_{\nu,\mu}=0; \,\, \vec{\nabla} \vec{J}_{\bar{\nu}} + \sum_{\mu} K_{\nu,\mu}=0; 
\label{const1} \\
&&\vec{\nabla} \vec{J}_\mu - \sum_{\nu} K_{\nu,\mu}=0; \,\, \vec{\nabla} \vec{J}_{\bar{\mu}} -  \sum_{\nu} K_{\nu,\mu}=0;
\label{const2} 
\eea
with $\vec{\nabla} \vec{J}_\nu\equiv \sum_{\hat{\alpha}} \nabla_{\hat{\alpha}} J_{\nu;i,\hat{\alpha}}$ being the lattice divergence and the vector notation $\vec{J}_\nu \equiv J_{\nu;i,\hat{\alpha}}$ along each link direction used. 
\begin{figure}[!htb]
\includegraphics[width=0.75 \columnwidth]{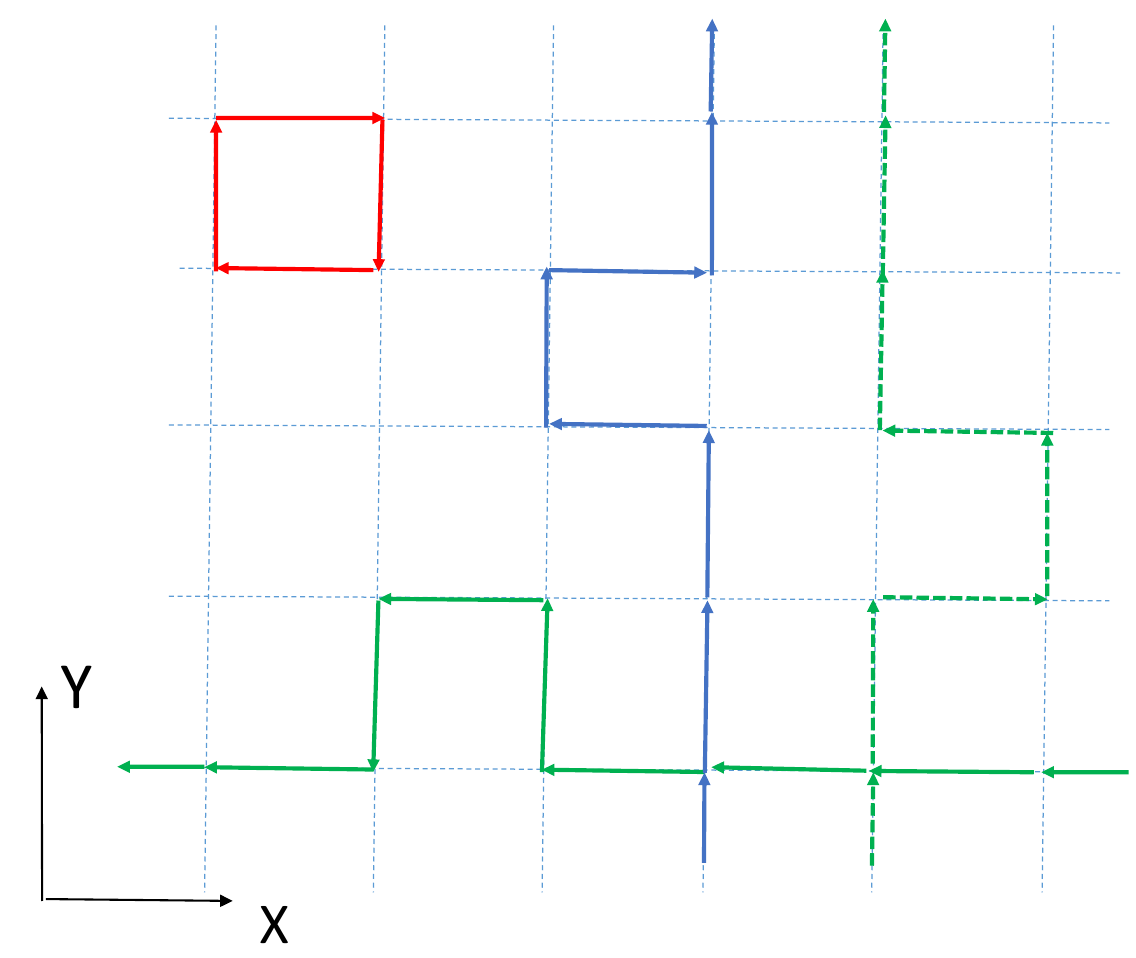}
\caption{ Typical configurations of the J-current link variables $\vec{J}_\nu$ in the case $g=0$ in Eqs.(\ref{HJ2}-\ref{const2}). Different colors indicate different components ($\nu$-values). Solid and dashed arrows of the same color indicate pairs $\vec{J}_\nu$ and $\vec{J}_{\bar{\nu}}$. Periodic boundary conditions are imposed along each direction, resulting in non-zero windings $W_{\nu,\hat{\alpha}}$. In the sketch shown $W_{1,\hat{y}}=1,\, W_{\bar{2},\hat{y}}=1, W_{2,\hat{x}}=-1$, where $\nu=1$ refers to blue, $\bar{2}$---to dashed green, $2$---to green. The red loop (say, $\nu=3$) does not create any winding.  }
\label{fig1}
\end{figure}
\begin{figure}[!htb]
\includegraphics[width=0.75 \columnwidth]{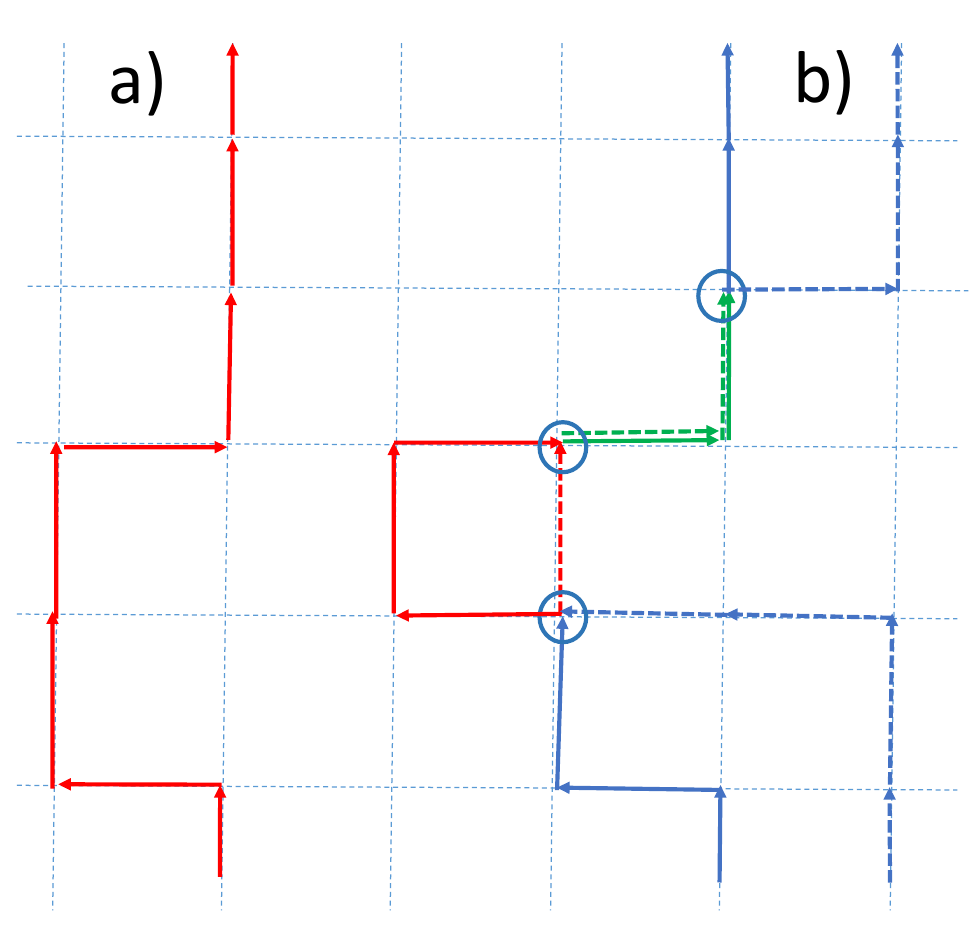}
\caption{ Typical configurations of the J-currents  (notations are the same as in Fig.~\ref{fig1}) in the case $g\neq 0$ in Eqs.(\ref{HJ2}--\ref{const2}). a) A loop characterized by the winding $W_{\nu,\hat{y}}=+1$; b) The double loop where Kirchhoff's rule is modified so that a pair of currents $\vec{J}_\nu,\, \vec{J}_{\bar{\nu}}$ transforms into a pair $\vec{J}_\mu,\, \vec{J}_{\bar{\mu}}$ with $\mu \neq \nu$   
The circles indicate sites where $K_{\nu,\mu}\neq 0$. This double loop contributes the winding $\sum^{N_v}_{\nu=1}W_{\nu, \hat{y}}=2$. In the PSF phase only such double loops proliferate.  }
\label{fig2}
\end{figure}

In terms of these variables, the stiffnesses (\ref{stiff}) become
\bea
\rho_{\Phi,\hat{\alpha}} &=&\frac{1}{L^{d-2}}  \langle \left(\sum^{N_v}_{\nu=1}W_{\nu, \hat{\alpha}}\right)^2\rangle,
\label{Wp} \\
\rho_{\nu,\hat{\alpha}} &=&\frac{1}{L^{d-2}}  \langle \left(W_{\nu, \hat{\alpha}} - W_{\bar{\nu}, \hat{\alpha}}\right)^2\rangle, 
\label{Wnu}
\eea
where $ \nu=1,2,...,N_v/2$ in Eq.(\ref{Wnu}), and the definition
\beq
W_{\nu,\hat{\alpha}}= \frac{1}{L}\sum_{i} J_{\nu;i,\hat{\alpha}}
\label{W}
\eeq
is used.  This definition of the stiffnesses in terms of the winding numbers is analogous to that introduced in Ref.\cite{Ceperley}.

If $g=0$, the only admissible values of $K_{\mu,\nu, i}$ are zeros. In this case, the system features $N_v$ independent condensates, with $N_v$ link variables $J_\nu$ forming closed oriented loops, that is, obeying Kirchhoff's current conservation law on each site $\vec{\nabla} \vec{J}_\nu=0$. Examples of such closed loops are shown in Fig.~\ref{fig1} featuring four types of loops, with three of them winding over the whole sample. 
In this case,
the quantities $W_{\nu,\hat{\alpha}}$ (\ref{W}) become windings numbers of the loops.
More specifically, the definition (\ref{W}) in the case $g=0$ is equivalent to  $W_{\nu,\hat{\alpha}}= \sum_{i\in S } J_{\nu;i,\hat{\alpha}}$ where the summation runs over a cross section $S$ of a sample. Since the link-currents obey the Kirchhoff rule, $ W_{\nu,\hat{\alpha}}$ is independent of the choice of the cross section position for its chosen orientation $\hat{\alpha}$.  
Because of the statistical independence and the symmetry between the components, all averages $\langle W_{\nu,\hat{\alpha}}^2\rangle$ are equal to each other and $\langle W_{\nu,\hat{\alpha}}W_{\mu,\hat{\alpha}}\rangle=0$ for $\mu \neq \nu$.
Thus, in this case $\rho_{\Phi,\hat{\alpha}}=(N_v/2)  \rho_{\nu,\hat{\alpha}} $, or
\beq
\rho_{\Phi,\hat{\alpha}}= \sum^{N_v/2} _{\nu=1}\rho_{\nu,\hat{\alpha}} .
\label{NvSF}
\eeq  
This relation characterizes the   N$_v$-SF, where $e^{i\varphi_\nu}$ is condensed and robust for all $\nu$. 

if $g$ is finite, the Kirchhoff rule is violated for the J-currents at sites where $K_{\nu,\mu}\neq 0$ in Eqs.(\ref{const1},\ref{const2}). There a pair of the currents $\vec{J}_\nu$ and $\vec{J}_{\bar{\nu}}$ can transform into a pair $\vec{J}_\mu$ and $\vec{J}_{\bar{\mu}}$ with $\mu \neq \nu$. Examples of such transformations are shown in Fig.~\ref{fig2} at sites marked by open circles.   Accordingly, while each individual $W_{\nu,\hat{\alpha}}$ is not a winding number, the combinations $W_{+,\hat{\alpha}}=\sum^{N_v}_{\nu=1}W_{\nu, \hat{\alpha}}$ and $W_{-,\nu, \hat{\alpha}}=W_{\nu, \hat{\alpha}} - W_{\bar{\nu}, \hat{\alpha}}$ are. 
Then, in the PSF phase loops with finite $W_+$ proliferate (see the type b) in Fig.~\ref{fig2}) , while the loops with finite $W_-$ (that is, the type a) in Fig.~\ref{fig2}) do not.  The N$_v$-SF is characterized by the proliferation of both types of loops $W_+$ and $W_-$.  Jumping ahead, it is worth mentioning that the relation (\ref{NvSF}) has been found to hold with high accuracy in the N$_v$-SF phase at finite $g$, that is, as though the terms $\sim g$ in Eq. (\ref{HJ}) become irrelevant. 

\subsection{Results of simulations in 3D.}
\begin{figure}[!htb]
\includegraphics[width=0.95 \columnwidth]{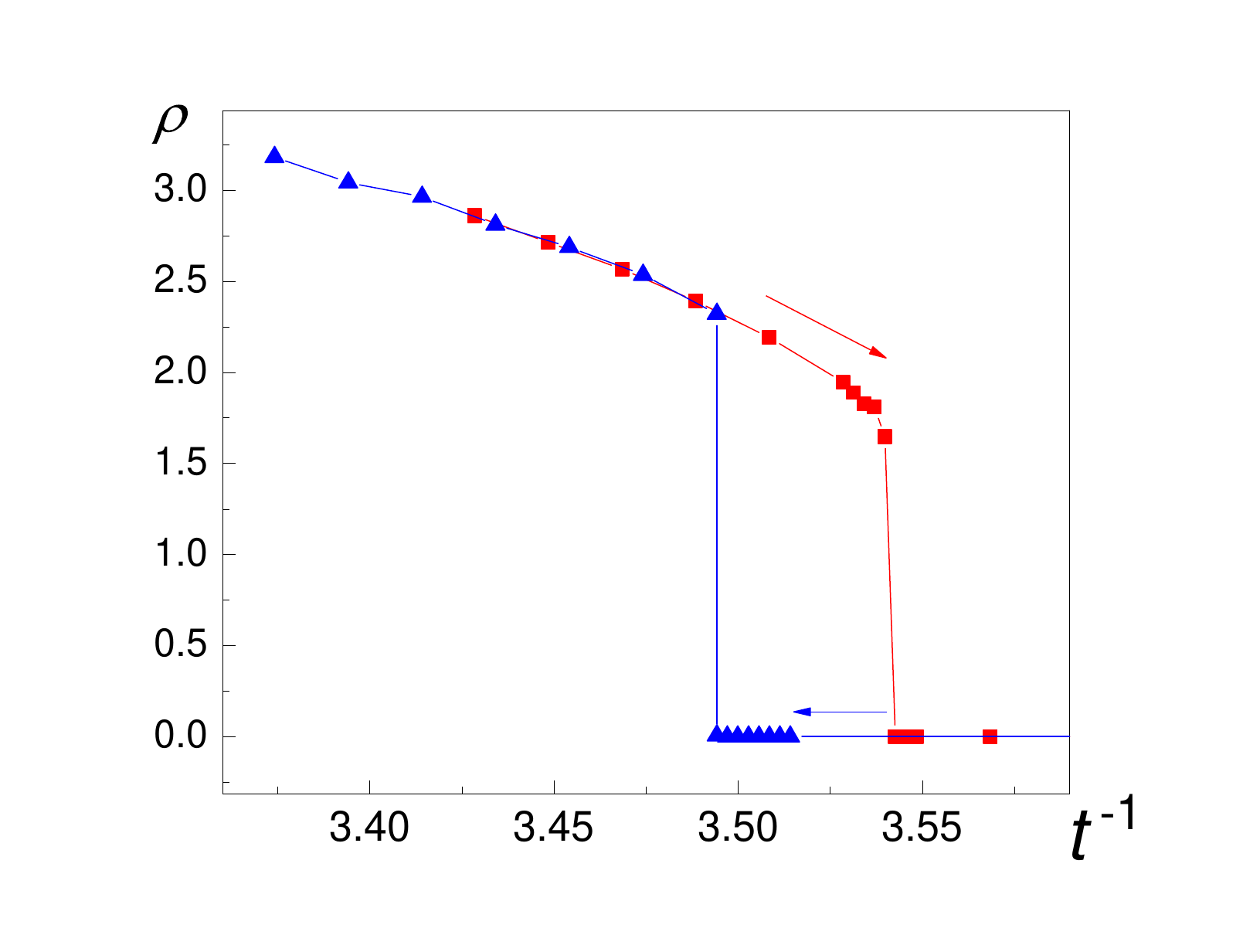}
\caption{The hysteretic behavior of $\rho = \sum_{\hat{\alpha}} \rho_{\Phi,\hat{\alpha}}\approx \sum_{\hat{\alpha}} \sum^{N_v/2}_{\nu=1} \rho_{\nu,\hat{\alpha}}$ in a 3D sample, $L=32$, $N_v=6$.    }
\label{fig4}
\end{figure}

Simulations have been conducted by utilizing the Worm Algorithm \cite{WA} modified for the case of double loops \cite{2WA} exemplified in Fig.~\ref{fig2}. It was found that changing $g\geq 1$ weakly affects the results. Accordingly, the limit $g=\infty$ has been considered in the action (\ref{HJ2}) , which leaves the system with just one parameter, that is, $t$. In 3D case, the pairing fluctuations are driving the system into a Ist order phase transition for $N_v\geq 4$. Fig.~\ref{fig4} shows a strong hysteresis as the system
heats up from the N$_v$-SF and cools down from the normal phase.  Such a behavior is a telltale sign of the discontinuous transition. It is also instructive to note that the condition (\ref{NvSF}) is fulfilled within the statistical errors of a couple of percents which implies that the pairing correlations do not lead to the paired phase, and, instead,---to the Ist order condensation transition.
  
\begin{figure}[!htb]
\includegraphics[width=0.95 \columnwidth]{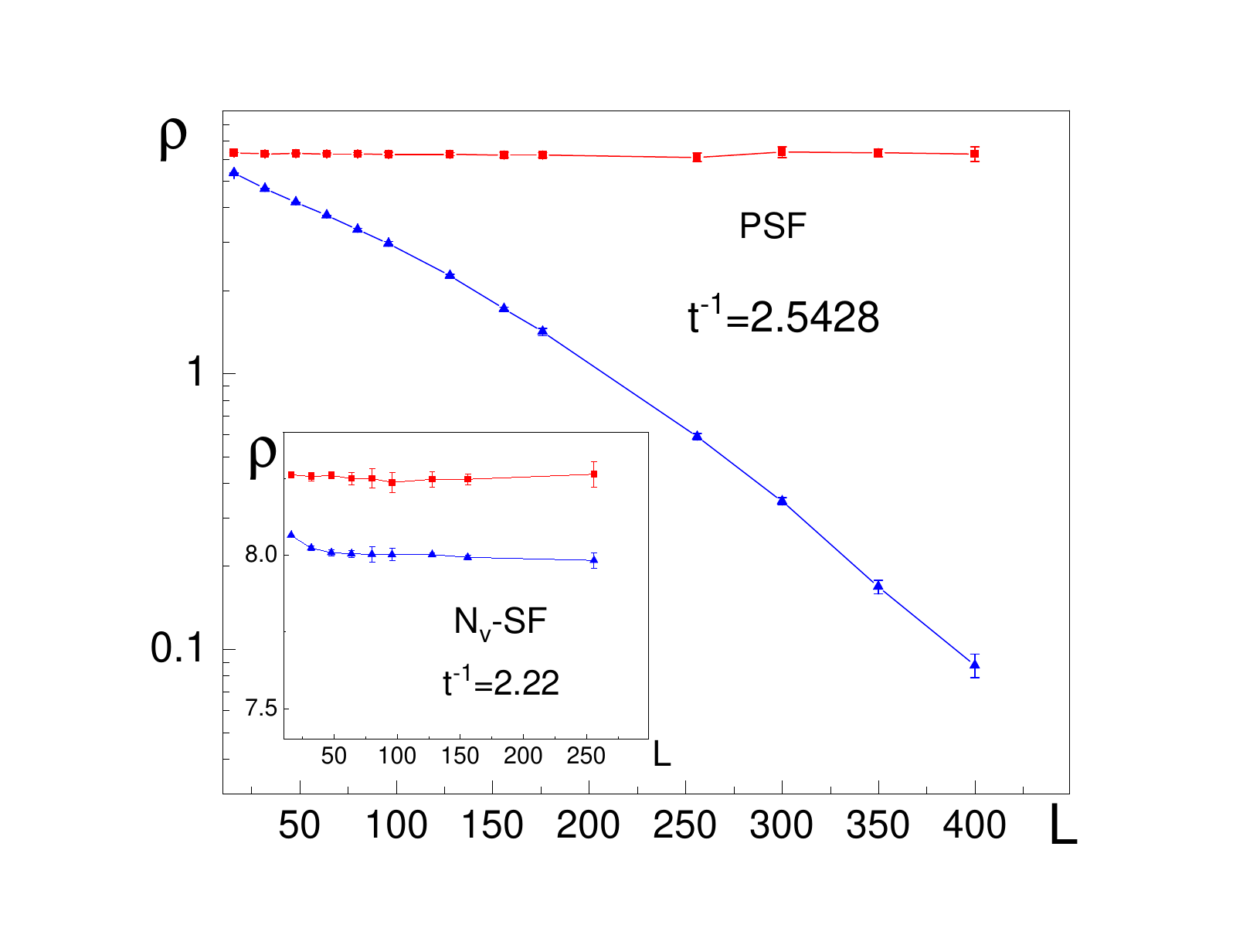}
\caption{2D stiffnesses $\rho_\Phi$ (squares), Eq.(\ref{Wp}), and $\sum_\nu \rho_\nu$ (triangles), Eq.(\ref{Wnu}), vs  linear system size $L$ for two values of the temperature $t^{-1}$ shown, $ N_v=10$. Main panel: In the PSF, while $\rho_\Phi$ is, practically, size independent, $\rho_\nu$ falls down with $L$;
Inset: The same quantities for smaller $t^{-1}$, that is, in the N$_v$-SF phase, where both stiffnesses saturate to size-independent values which are close to each other within 3\%.     }
\label{fig5}
\end{figure}
\begin{figure}[!htb]
\includegraphics[width=0.95 \columnwidth]{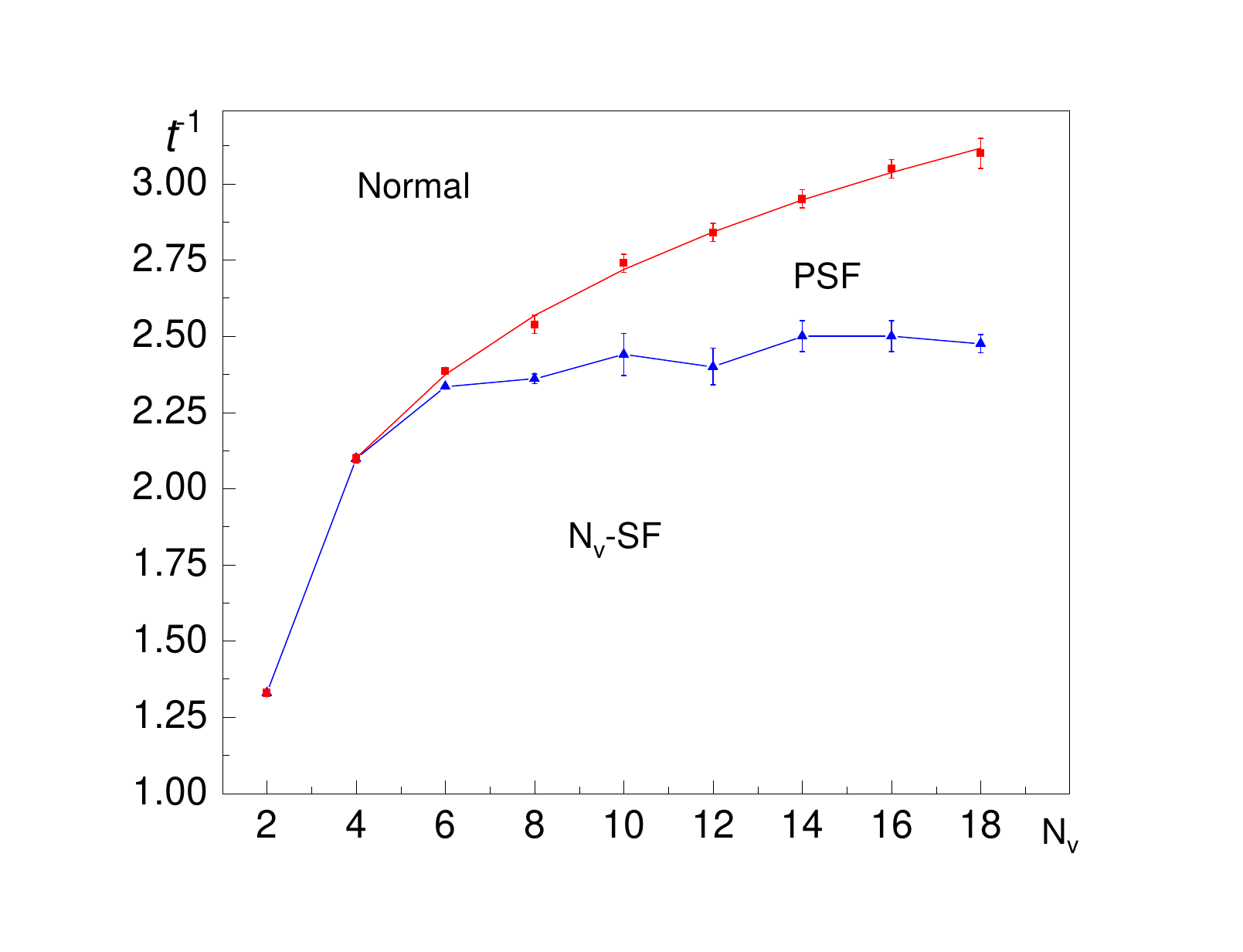}
\caption{The phase diagram for the model (\ref{Htot}) in the Villain approximation. The line in the upper data set is the fit by $t^{-1}= a \ln N_v +b$ with $a=0.68 \pm 0.01,\,\, b=1.16 \pm 0.02$.   }
\label{fig6}
\end{figure}
\begin{figure}[!htb]
\includegraphics[width=0.95 \columnwidth]{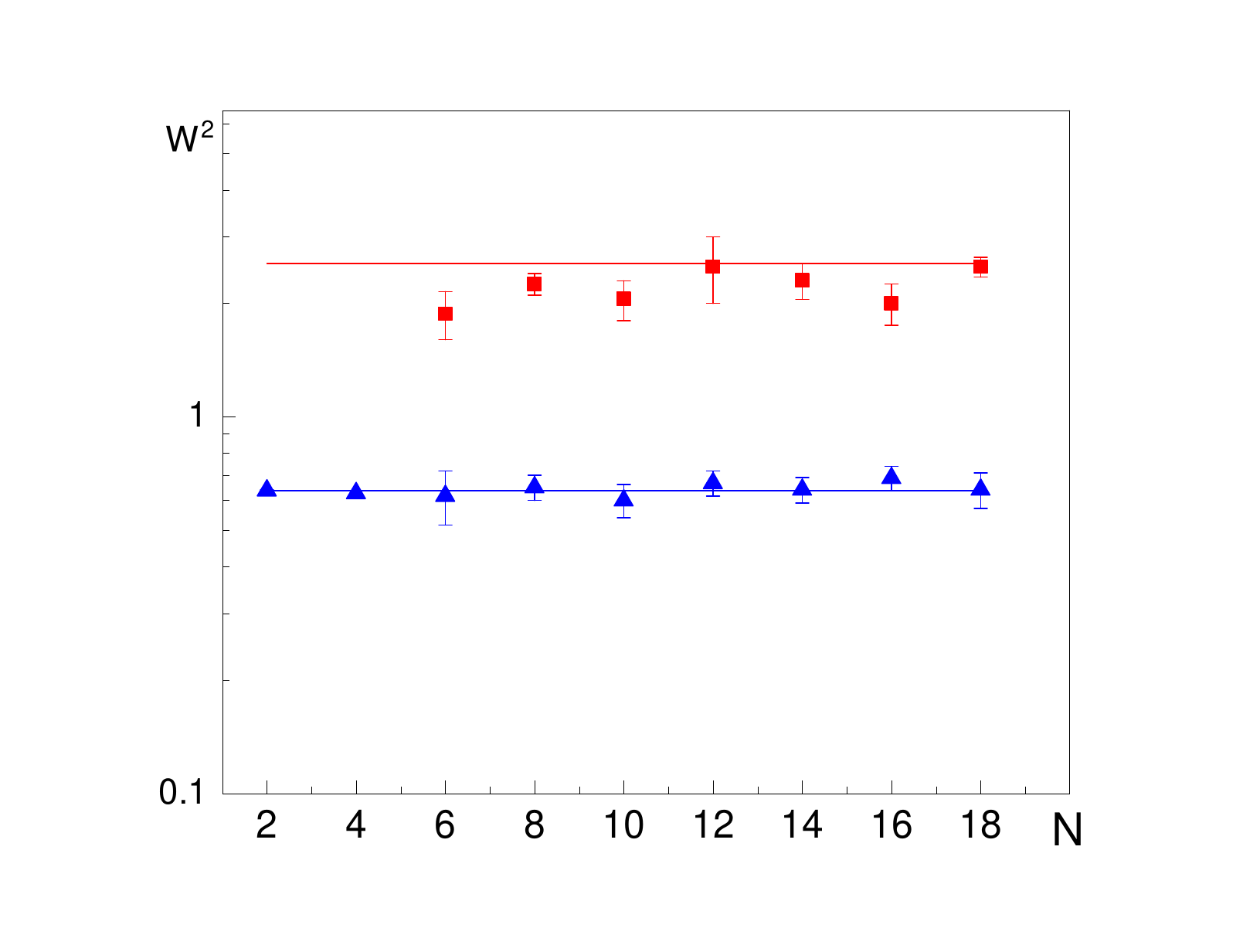}
\caption{ Universal jumps of the stiffnesses vs $N_v$ at the phase lines of the phase diagram, Fig.~\ref{fig6}. The lower and the upper straight lines correspond, respectively, to the values $2/\pi$ and to $8/\pi$.   }
\label{fig7}
\end{figure}
\begin{figure}[!htb]
\includegraphics[width=0.75 \columnwidth]{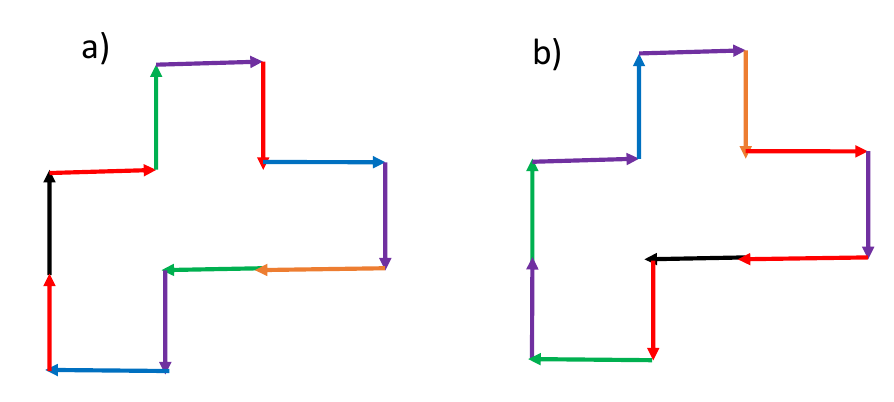}
\caption{ Double loops consisting of pairs of segments of the J-currents $J_\nu, \,\, J_{\bar{\nu}} $ with various $\nu$. [Each colored arrow represents a pair of the J-currents $J_\nu, \,\, J_{\bar{\nu}} $]. The additional entropy is due to swapping positions of the segments. The loops a) and b) have the same shape and different sequence of the double segments.}
\label{fig8}
\end{figure}

\subsection{Results of simulations in 2D}

The existence of the PSF and N$_v$-SF phases in 2D case is demonstrated in Fig.~\ref{fig5}. While $\sum_{\hat{\alpha}}\rho_{\Phi, \hat{\alpha}}$ as a function of $L$ saturates to a size-independent value in the PSF, the stiffness $\sum_{\hat{\alpha}}\rho_{\nu, \hat{\alpha}}$ falls to zero in the TD limit. In the N$_v$-SF phase both stiffnesses saturate to constant values which are close to each other within few percents. 

The phase diagram for different values of $N_v$ is shown in Fig.~\ref{fig6}. The paired phase occurs first upon lowering the temperature $t^{-1}$ for $N_v\geq 6$. Further lowering the temperature leads to a transition from the PSF to the N$_v$-SF. Both transitions belong to the class of the Berezinskii-Kosterlitz-Thouless (BKT) transition which can be characterized by depairing of vortices with the lowest circulation of $\Phi$ at the upper boundary and of $\phi_\nu$ at the lower one. 
Accordingly, the superfluid stiffnesses (\ref{Wp}) and (\ref{Wnu}) undergo the universal jumps  $8/\pi$ and $2/\pi$, respectively. These values have been found by  fitting the dependencies of $\rho_\nu$ and $\rho_\Phi$ on $L$ by the separatrix of the renormalization group solution similarly to the method used in Ref.\cite{borom}. The results of such fits are presented in Fig.~\ref{fig7}. The comment is in order on why the first value is 4 times larger.
The universal value of $\langle W_{\nu,\hat{\alpha}}^2\rangle =2/\pi$ implies that at the transition the statistics of the winding numbers $W$ is dominated by a single macroscopic loop with $W_{\nu,\hat{\alpha}}=\pm 1$ of the type shown in a), Fig.~\ref{fig2}. In the case of the pairing phase, such loops are characterized by $\sum_\nu W_{\nu,\hat{\alpha}} =\pm 2$, as exemplified in b) of Fig.~\ref{fig2}. Accordingly, the universal value is factor 4 larger. This observation is another indication of the emergence of the PSF.  

\subsection{DNA analogy and the perspective for n-compound phases.}

The dual formulation provides a natural way to understand why the width of the PSF phase as a function of $N_v$, Fig.~\ref{fig6}, grows much slower than the linear dependence $t^{-1} \sim N_v$ for $N_v>>1$  following from the simplistic analytical argument given above in Sec.\ref{SEC1}. The key is the balance of the energy $\approx  L/t $ of the double loop (of the type b)) sketched in Fig.~\ref{fig2} of length $L$ and the entropy resulting from the shuffling of the J-current pairs $\nu, \bar{\nu}$ along the length of the loop as depicted Fig.~\ref{fig8}, which is a simplified version of the double loop shown in b) of Fig.~\ref{fig2}.This brings up an analogy to the double helix "DNA" composed of $N_v$ complimentary bases. Each lattice link can carry a pair of the currents $J_\nu,\,\, J_{\bar{\nu}}$ with equal probability.   Accordingly, the contribution to the partition function of a double loop can be estimated as $\sim \exp( -L/t) (N_v/2)^L$. Thus, for 
\beq
t^{-1}< t^{-1}_2 \approx \ln(N_v/2),
\label{LN}
\eeq
the free energy as a function of $L$ becomes negative, implying that the double loop proliferates. It is worth noting that the fit line of the upper boundary in Fig.~\ref{fig6} gives $t^{-1}_c \approx 0.68 \ln(N_v) $.
In the estimate (\ref{LN}) no  contribution from the shape fluctuations is taken into account. 
The same argument allows understanding why the lower boundary in Fig.~\ref{fig6} is, practically, independent of $N_v>>1$.
At this boundary single loops of the type a), Fig.~\ref{fig2}, proliferate. Accordingly, the only entropic contribution to such a loop comes from its shape fluctuations. 

The DNA-analogy opens up a clear perspective on the role of internal Josephson couplings of the orders $n$ higher than $n=2$ described by the term (\ref{H_J}). It is possible to argue that such orders are irrelevant if $g$ in Eq.(\ref{H_J}) is finite. That is, no $n$-compound phase (characterized by the order parameter $\sim \Psi_{\nu_1}\Psi_{\nu_2}...\Psi_{\nu_n}=\psi_{\nu_1}\psi_{\nu_2}...\psi_{\nu_n} $) with $n>2$ should exist.
The $n$-th order internal Josephson effect can occur between groups of valleys characterized by $\sum^n_{\nu=1} \vec{Q}_\nu=0$. 
For example, in the case $N_v=6$, Fig.~\ref{fig0},  two such groups exist for $n=3$---consisting of the valleys $(1,3,\bar{2})$ and $(2,\bar{1},\bar{3})$.
In general, there are $N_v/n$ of such groups. Then, the "DNA"-argument used to obtain the estimate (\ref{LN}) applies to the case of arbitrary $n$ as well. 
In other words, the formation of the n-compound phase requires proliferation of a loop each element of which consists of $n$ elementary J-currents of different "colors" forming the groups. Then, the contribution to the partition function of the loop of length $L$ consists of its energy $\sim nL/2t$ and the multiplicity $(N_v/n)^L$ due to swapping of the groups of elementary J-currents. Thus, the proliferation should occur for $t^{-1}< t^{-1}_n \approx \frac{2}{n}\ln(N_v/n)$, which lies below the boundary (\ref{LN}) for $n>2$ unless accidentally $g=0$ in (\ref{H_J}).

\section{Detection}
The question about detecting dark excitonic condensate was addressed many times in the past with the focus on the coherence of one-exciton density matrix. In respect to the system discussed here, a special attention should be paid to the method discussed in Ref.\cite{Combescot1} and focused on the spin-controlled bright-dark exciton conversion. 
The term similar to (\ref{HJ1}) (cf. \cite{Combescot1}) can be written for the interconversion between pairs of the dark and bright excitons as
\beq
H_{db} = -\frac{\tilde{g}}{2} \int d^2x \sum_{\nu}  (\Psi^*_{b} \Psi^*_{b} \Psi_{\nu} \Psi_{\bar{\nu}} + c.c.), 
\label{HJ5}
\eeq
where $\Psi_b$ stands for the bright-exciton operator, and $\tilde{g}$ is some constant. In the PSF and N$_v$-SF the products $\tilde{\Psi}=\Psi_{\nu} \Psi_{\bar{\nu}}$ have their phases $\varphi_\nu$ all locked up into a single phase $\Phi$ which results in the constructive interference with respect to creating pairs of the bright excitons with opposite momenta. In this sense the term (\ref{HJ5}) does not allow distinguishing between PSF and N$_v$-SF phases of the dark condensate. However, the effect discussed in Ref.\cite{Adham} is sensitive to the coherence of $\Psi_\nu$ rather than to $\tilde{\Psi}$. Thus, in the PSF phase, while the two-exciton effect described by Eq.(\ref{HJ5}) occurs, the photon-exciton-phonon beats of the type \cite{Adham} will not take place.

Another option to be considered is with respect to the recently observed resonant upconversion of two dark excitons into a single bright exciton \cite{up}.  It has been conjectured that this effect might be related to the spontaneous coherence of the dark excitons \cite{up}. However, this effect too cannot distinguish between the PSF and N$_v$-SF phases. Accordingly, it should also be considered together with the effect \cite{Adham}.

\section{Discussion} 
 The proposed excitonic intervalley pairing is induced by purely thermal fluctuations. Indeed, as temperature approaches zero, the correlations between windings $W_\nu$ fade away leading to practical independence  of the components which can be seen in the fulfillment of the condition (\ref{NvSF}) as $t^{-1} \to 0$. In other words, there is no explicit attractive force between excitons causing pairing. This effect is another example of the thermally induced pairing (cf. Ref.\cite{Egor}). An open question is how quantum fluctuations may change this situation. In view of the quantum-to-classical correspondence, a 2D quantum  system is equivalent to a 3D classical, with its third dimension playing the role of the imaginary time in the range from 0 to the inverse temperature $\beta$. Thus, it is natural to assume that the quantum fluctuations in a 2D system will induce the first order phase transition between normal state and N$_v$-SF as $\beta \to \infty$. In other words, no PSF should exist at $T=0$. This phase is expected to exist only above some critical temperature $T^*$. In this sense, quantum fluctuations appear to be detrimental to the pairing effect. Detailed numerical studies of the quantum to classical crossover in the the system of intervalley excitons is a subject of a future work.

The PSF phase has been discussed here in the context of intervalley excitons in layered systems. To what extent the effect can be realized in other materials is an open question. An interesting possibility for realizing the moat (that is,  $N_v =\infty $) dispersion in a driven cold atomic system has been suggested in Ref. \cite{Sedrakyan4}. To what extent this method allows obtaining the discrete version of the moat (that is, finite $N_v$) remains to be seen.

{\it Acknowledgments}.---This work was supported
by the National Science Foundation under the grant DMR-2335905. Useful discussions with Boris Svistunov and Nikolay Prokof'ev are appreciated.

\end{document}